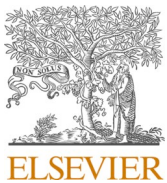
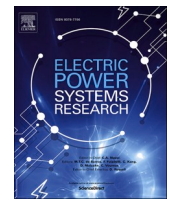

# Experimental validation of ultra-shortened 3D finite element electromagnetic modeling of three-core armored cables at power frequency

Juan Carlos del-Pino-López [*], Pedro Cruz-Romero

*Department of Electrical Engineering, Universidad de Sevilla, Camino de los Descubrimientos s/n, Sevilla 41092, Spain*

ABSTRACT

Due to recent advances, the numerical analysis of submarine three-core armored cables can nowadays be developed through the finite element method (FEM) in a small slice of the cable. This strongly reduces the computational burden and simulation time. However, the performance of this ultra-shortened 3D-FEM model is still to be fully assessed with experimental measurements. This paper focuses on this validation for an extensive variety of situations through the experimental measurements available in the specialized literature for up to 10 actual cables. In particular, it deals not only with relevant calculations at power frequency, like the series resistance and inductive reactance or the induced sheath current, but also with other aspects never analyzed before through 3D-FEM simulations, such as the zero sequence impedance, the magnetic field distribution around the power cable, as well as side effects due to the nonlinear properties of the armor wires. All this considering different armoring and sheath bonding configurations. Results show a very good agreement between measured and computed values, presenting the ultra-shortened 3D-FEM model as a suitable tool for the analysis and design of three-core armored cables, and opening the possibility to reduce the need of extensive experimental tests in the design stage of new cables.

## 1. Introduction

The number of offshore wind power plants (OWPPs) has significantly increased during the last decades (e.g., the cumulative installed capacity in Europe has raised from 2 GW to 22 GW in 10 years [1]), and this trend will be increasing in the following years [1]. This situation is leading to larger OWPPs (usually placed far from the shore), forcing the technology to new limits, especially regarding three-core export power cables [2]. In this sense, it is well-known that the IEC 60287 standard [3] leads to oversized three-core armored cables (TCACs), so it is becoming an urgent need to have appropriate tools for designing and evaluating the performance of this type of cable.

During the last decade, new advances have been developed in this topic area, where the use of numerical simulations, such as the finite element method (FEM), has become one of the most powerful and versatile tools for analyzing TCACs. It was initially employed to evaluate the power losses in TCACs, as in [4–6], where the use of additional constraints in 2D-FEM models were proposed to take into account the relative twisting between the armor wires and the phase conductors. This feature has led to quite a few studies aiming at understanding and characterizing phenomena like induced losses [7–9], thermal behavior [10–14], and electrical parameters [15–19], as well as paving the way for the development of new analytical approaches to improve the IEC standard [20–22].

However, although 2D-FEM simulations are more accurate than the IEC standard [5–7,9], the longitudinal component of the magnetic field is omitted, so 3D geometries are still required for properly evaluating the influence of the armor and three-core twisting. A first 3D-FEM approach was developed in [4,22–24], where 3D simplified geometries were employed. More recently, a 3-meter long 3D-FEM model, more accurate to reality, was presented in [25] for the first time in the literature, showing the benefits of using 3D geometries instead of 2D ones, as well as the deviations observed in the results provided by [3]. However, huge computational resources were required. To overcome this issue, different simplifying assumptions have been proposed in the literature, as in [26,27], where the length of the 3D geometry is shortened until end effects start to influence the results. Alternatively, other authors suggest the use of a coarser mesh [28].

Nevertheless, despite these remarkable efforts, the computational requirements and the simulation time were still far from those observed in 2D-FEM simulations. In this sense, the use of rotated periodicity as an

---

## Nomenclature

| | |
|---|---|
| $B_h$ | horizontal component of the magnetic field induction (μT) |
| $B_l$ | longitudinal component of the magnetic field induction (μT) |
| $B_v$ | vertical component of the magnetic field induction (μT) |
| $D_a$ | armor outer diameter (mm) |
| $d_a$ | armor wire diameter (mm) |
| $d_c$ | conductor diameter (mm) |
| $D_s$ | sheath outer diameter (mm) |
| $D_{core}$ | power cores outer diameter (mm) |
| $d_{sea}$ | Carson's depth (m) |
| $e_s$ | sheath thickness (mm) |
| $f$ | frequency (Hz) |
| $I_a$ | induced armor current (A) |
| $I_{max}$ | rated current (A) |
| $\vec{J_e}$ | external current density (A/m$^2$) |
| $L$ | length of the 3D geometry (m) |
| $L'$ | length of the 3D geometry when $N' < N$ (m) |
| $L_a$ | armor lay length (m) |
| $L_c$ | phase conductor lay length (m) |
| $N$ | total number of armor wires |
| $N'$ | number of steel armor wires ($N' \leq N$) |
| $R_0$ | zero sequence resistance (Ω/km) |
| $S_n$ | cable cross section (mm$^2$) |
| $T_{amb}$ | ambient temperature (°C) |
| $V_a$ | induced armor voltage (V) |
| $V_r$ | rated voltage (kV) |
| $X_0$ | zero sequence reactance (Ω/km) |
| $\theta$ | rotational displacement (rad) |
| $\theta'$ | rotational displacement when $N' < N$ (rad) |
| $\sigma_c$ | conductor electrical conductivity (MS/m) |
| $\sigma_s$ | sheath electrical conductivity (MS/m) |
| $\sigma_a$ | armor electrical conductivity (MS/m) |
| $\mu_r$ | relative permeability |
| $\mu_r'$ | real part of the complex permeability |
| $\mu_r''$ | imaginary part of the complex permeability |
| $\rho_{sea}$ | sea electrical resistivity (mΩ·m) |
| $\omega$ | angular frequency (rad/s) |

additional boundary condition was proposed in [29] for reducing the length of the 3D geometry to be simulated, being shorter than 2 m in most situations. Subsequent studies [30,31] proposed further improvements, reducing the length of the 3D-FEM model to the maximum in [32–35], with values as short as 2-3 cm. This ultra-shortened 3D-FEM model (USM) reduces simulation time below 1 minute without loss in accuracy compared to larger 3D-FEM models [34].

Despite the improvements mentioned above, the USM remains to be thoroughly and systematically confronted with experimental measurements, not only for evaluating its performance regarding accuracy of results, but also to determine to what extent this tool can be valid for the analysis of other features and phenomena of TCACs. In this sense, few studies have compared 3D-FEM simulations with experimental measurements in a limited number of cables [27–29], mainly due to the difficulties for the academia to have access to a piece of cable for testing purposes, since they are specifically built on a project-by-project basis. Thus, besides the laboratory requirements, experimental tests are only possible on actual installations [36–38] or in collaboration with cable manufacturers, as can be derived from the specialized literature [4–6,9, 22,24,27,28,39–45]. Consequently, simulations tools, like USM, provide a good alternative for supplementing costly experimental setups, making easier the analysis and design of TCACs.

Having all this in mind, this paper develops the experimental validation of the USM presented in [32–35]. To this aim, a faithful replication of previously published experimental studies on a set 10 actual TCACs, representing a wide variety of voltages, cross sections and conductor materials, has been performed. Power-frequency electromagnetic analyses are performed for a wide range of situations, evaluating the relative differences between measurements and simulated values regarding the series resistance, inductive reactance and the induced sheath current. Also, other aspects never analyzed before through 3D-FEM simulations are assessed, such as the zero sequence impedance, the distribution of the magnetic field induction (MF) around TCACs, as well as the presence of a third harmonic in the induced armor voltage due to the nonlinear properties of the armor wires. All this is performed for different armoring and bonding configurations. Be aware that loss allocation in conductor, sheath and armor are not addressed in this study since they can not be accurately obtained just from experimental measurements, as shown in [31,39–41]. Results highlight a good matching between measurements and simulated values, becoming, with certain simulation cautions, a powerful and cost-effective tool to analyze the electromagnetic behavior of TCACs with similar accuracy than experimental tests.

## 2. Ultra-shortened 3D-FEM model

As described in [33,34], the USM makes use of rotated periodicity for reducing the length of the 3D geometry to be simulated up to

$$L = \frac{1}{N \cdot \left| \frac{1}{L_c} \pm \frac{1}{L_a} \right|}, \quad (1)$$

where + is taken when the armor wires and the phase conductors are twisted in different directions (contralay), and - if twisted in the same direction (unilay). Thus, rotated periodicity boundary conditions are applied by mapping the source boundary into the destination boundary, where the rotational displacement of the power cores $\theta$ must be considered to get a perfect matching between both boundaries (Fig. 1a), defined as

$$\theta = \frac{2\pi L}{L_c}. \quad (2)$$

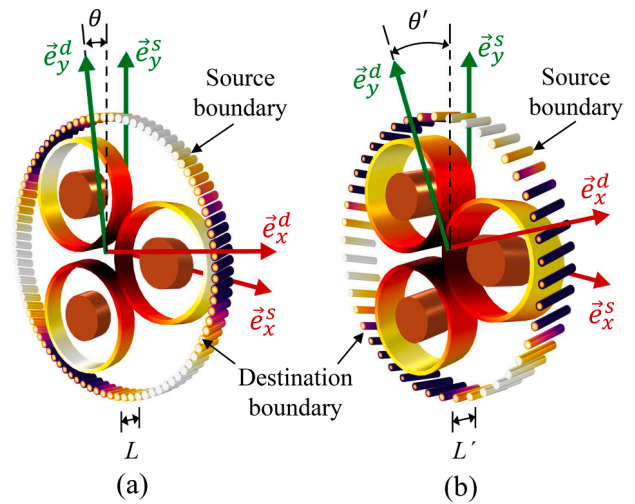

**Fig. 1.** Rotated periodicity in (a) fully armored ($N$ steel armor wires) and (b) half-armored cables ($N' \approx N/2$ steel armor wires).

The USM is implemented in COMSOL Multiphysics® for frequency-domain analyses by means of the AC/DC module [46,47]. The "Coil" feature included in COMSOL is employed to set the three-phase currents in the conductors, as well as the sheath and armor bonding (sheath/armor currents set to 0 A for the single point configuration (SP), and the voltage to 0 V in the solid bonding case (SB)).

This results in a simulation time below 1 minute without loss in accuracy compared to larger 3D-FEM models [34]. Consequently, finer meshes and more complex models can be simulated without greatly increasing the computational requirements [35]. Moreover, (1) and (2) are applicable for any value of $N$, including the cases when some of the steel wires are removed ($N' < N$) or replaced by polyethylene (PE) separators ($N' \approx N/2$, named here as half-armored cables). In these cases, the computation time increases due to the larger length of the resulting 3D model, ($L' > L$ in Fig. 1b), but it still remains below 2 minutes. Eventually, it should be pointed out that this approach is also applicable to time-domain simulations, as it will shown later in this work, although simulation time increases noticeably depending on the time stepping.

On the other hand, special attention must be devoted to the size of the simulation domain, especially when computing the zero sequence impedance or the MF distribution around TCACs, since the outer boundary must be far enough for obtaining accurate results. Customarily, this distance can be reduced in any FEM software by applying a nonlinear coordinate transformation to the surrounding medium layer, having the effect of stretching it to almost infinity. This is done in COMSOL Multiphysics through the "infinite element domain" feature [46]. Although this increases the number of unknowns, the computation time is barely affected. Further details will be commented in the following sections.

Finally, the armor $\mu_r$ can be modeled as nonlinear, including hysteresis losses through a complex relative permeability ($\mu_r' - j\mu_r''$) [46,47]. All this influences the simulation time, which increases in about 50 %. However, due to the reduced 3D geometry, it remains always below 4 minutes for the worst case (half-armored cables).

## 3. Case studies

As commented earlier, a number of experimental studies can be found in the literature, although many of them do not provide enough data for being replicated in 3D-FEM simulations. Table 1 summarizes the main dimensions and properties of the most detailed study cases found in the literature. It must be pointed out that just a few geometrical or material parameters were missing in some of the references, so that it was still feasible in practical terms to estimate them by comparison with similar TCACs or to derive them from other data provided by the authors.

As can be seen, Table 1 includes lead sheathed cables with different cross sections, voltages, conductor materials, armor twisting (contralay (cont.) and unilay (uni.)), as well as magnetic and non-magnetic ($\mu_r = 1$) armors. It should be noted that the case of a half-armored cable is also included through cable C7.

Regarding the material properties, references usually provide $\sigma_c$, $\sigma_s$ and $\sigma_a$ at their temperature during tests, being in most of the cases $T_{amb}$, since these are usually performed during short periods to avoid thermal effects. On the other hand, it should be pointed that, for cables C6 to C10, the value of $\mu_r$ for the armor wires was not provided in the references, so it was assumed to be made of a typical low grade (LG) steel like

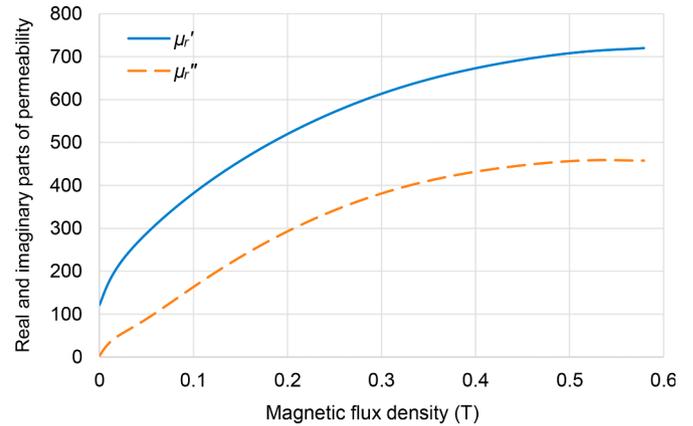

**Fig. 2.** Complex relative permeability for a LG steel.

**Table 1**
Main dimensions and properties of all the TCACs analyzed.

|  | C1a | C1b | C2a | C2b | C3 | C4 | C5 | C6 | C7 | C8 | C9 | C10 |
|---|---|---|---|---|---|---|---|---|---|---|---|---|
| Reference | [27,44] | [27,44] | [27,44] | [27,44] | [37] | [36,38] | [6] | [40] | [39] | [39] | [43] | [24] |
| $V_r$ (kV) | 245 | 245 | 245 | 245 | 150 | 220 | 132 | 220 | 132 | 115 | 30 | 132 |
| $I_{max}$ (A) | 1200 | 1200 | 1000 | 1000 | 650 | 675 | 732 | 500 | 722 | 507 | 200 | 800 |
| $S_n$ (mm$^2$) | 1600 | 1600 | 1200 | 1200 | 630 | 630 | 800 | 630 | 815 | 240 | 120 | 800 |
| Material | Cu | Cu | Al | Al | Cu | Cu | Cu | Cu | Al | Cu | Al | Cu |
| $d_c$ (mm) | 46.3 | 46.3 | 42.9 | 42.9 | 30.25 | 30.5 | 35 | 30.5 | 34.8 | 17.5 | 13.4 | 34.5 |
| $D_s$ (mm) | 104 | 104 | 99.5 | 99.5 | 80.6 | 92.1 | 87.6 | 92.1 | 76.8 | 56.1 | 37 | 86 |
| $e_s$ (mm) | 2.25 | 2.25 | 2.25 | 2.25 | 2.8 | 3 | 3.7 | 3 | 2.1 | 2 | 1.7 | 3.1 |
| $D_{core}$ (mm) | 109 | 109 | 104.5 | 104.5 | 85.6 | 97.3 | 92.4 | 97.3 | 80.8 | 60.1 | 41.57 | 89.72 |
| $D_a$ (mm) | 247.1 | 247.1 | 236.2 | 236.2 | 200.9 | 238.6 | 214.6 | 226.43 | 187.3 | 146.3 | 99 | 206 |
| $d_a$ (mm) | 5.6 | 5.6 | 5 | 5 | 6 | 5.6 | 5.6 | 5.6 | 4 | 5 | 4 | 6 |
| $N$ | 129 | 129 | 139 | 139 | 95 | 120 | 114 | 119 | 64 | 78 | 69 | 88 |
| $L_a$ (m) | 4 | 2.5 | 4 | 4 | 3.8 | 3 | 3.5 | 3 | 2.8 | 2.5 | 1.2 | 3.6 |
| $L_c$ (m) | 3.6 | 3.5 | 3.6 | 3.6 | 2.6 | 2.7 | 2.8 | 2.7 | 2.2 | 1.8 | 1 | 2.5 |
| Twist | cont. | uni. | cont. | cont. | cont. | cont. | cont. | cont. | cont. | cont. | cont. | cont. |
| $T_{amb}$ (°C) | 2 | 2 | 2 | 2 | 5 | 5 | 5 | 10 | 5 | 5 | 10 | 20 |
| $\sigma_c$ (MS/m) | 61.9 | 52.7 | 37.8 | 37.8 | 52.32 | 61.6 | 51 | 51.88 | 29.38 | 56.4 | 39.06 | 58.14 |
| $\sigma_s$ (MS/m) | 4.99 | 4.19 | 4.99 | 4.99 | 5.30 | 4.97 | 4.50 | 4.41 | 4.82 | 4.90 | 3.25 | 4.67 |
| $\sigma_a$ (MS/m) | 5.39 | 5.19 | 5.39 | 5.39 | 5.53 | 5.16 | 5.19 | 4.81 | 5.64 | 0.58 | 6.9 | 7.25 |
| $\mu_r$ | 150 − j50 | 150 − j50 | 150 − j50 | 1 | 300 | 300 | LG | LG | LG | LG | LG | LG |

the one employed in cable C5, characterized by the complex permeability shown in Fig. 2 [6,22].

In this work, the cables of Table 1 are simulated under the same operating conditions described in the references, which include the armor removal in some cases, leading to different combinations of armored and unarmored cables with sheaths in SP or SB configurations. In addition, some references analyze the case where half of the armor wires are removed. This is the case for cables C1a, C2a and C10, resulting in new cases hereinafter designed as C1a/2 ($N' = 65$), C2a/2 ($N' = 70$) and C10/2 ($N' = 44$). Therefore, the number of cases analyzed is higher than those presented in Table 1.

## 4. Simulation vs experimental results

In the following sections, the performance of the USM is assessed at power frequency (50 Hz) through the experimental results found in the references (detailed information regarding the experimental setups can be found in the references).

### 4.1. Series resistance

The equivalent series resistance of TCACs is experimentally obtained by measuring the phase current and the total real power involved during the test. A similar method is numerically applied in the USM. The comparison between measured and computed values for this parameter, including their relative difference, is summarized in Fig. 3 for the case of unarmored cables (labeled as NA), and in Fig. 4 for armored and half-armored ones. In both cases, sheaths are considered either in SP or SB.

As can be seen, the relative difference between measurements and simulation values is usually below 3% for the unarmored cases, while it slightly increases for the armored ones (below 6%). It is also observed that the greatest differences are usually related to SP cases, especially in armored cables. This may be caused by unbalanced supplied currents during laboratory tests, as reported by some authors [6]. Regarding the half-armored cases, results are also remarkable, with differences below 2%, except for cable C10/2, although it is still below 5%.

Another aspect that may influence the series resistance is the magnitude of the supplied current. Different studies suggest that the nonlinear properties of the armor wires lead to higher resistances as long as the supplied current increases. This is corroborated by Fig. 5a, where the calculations and measurements available for 4 cables with LG steel armor and 1 unarmored cable (C10 NA SB) are depicted. As can be observed, except for the unarmored case, the series resistance increases with the current in both, measurements and simulation results, with a relative difference below 4 % in all the cases (Fig. 5b).

It should be remarked that, for the sake of simplicity, the phase current has been normalized (p.u.) with the maximum current employed during each test, so that the same horizontal scale can be employed for all the cables represented in Fig. 5.

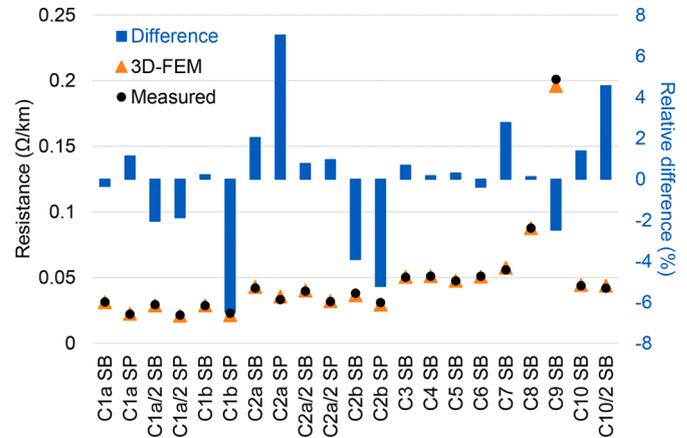

Fig. 4. Armored cables (SB and SP): measured and calculated series resistance (markers) and relative differences (bars).

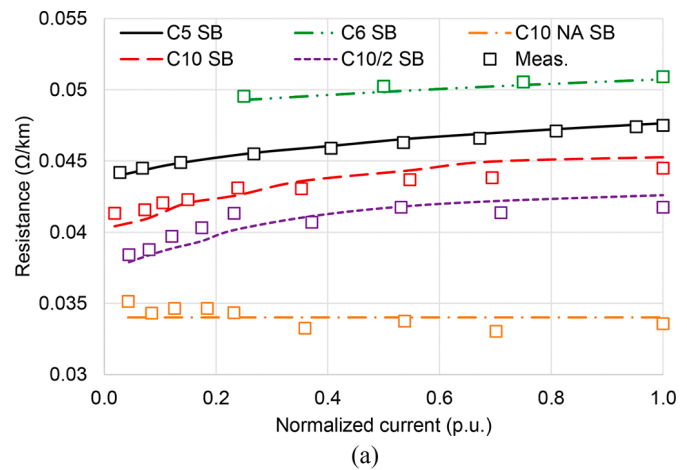

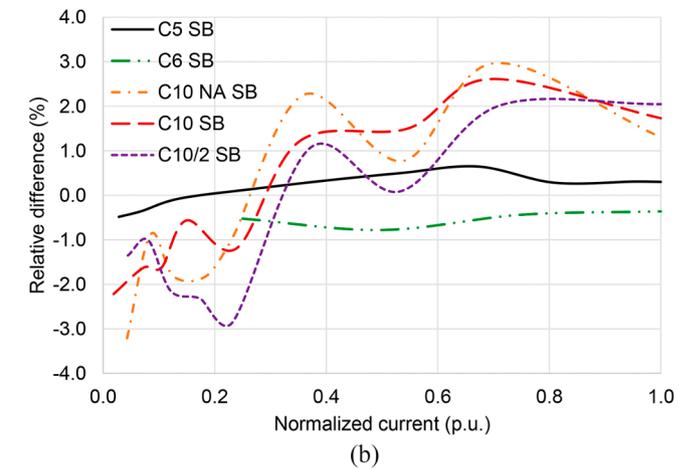

Fig. 5. (a) Evolution of the measured and calculated series resistance with phase current and (b) relative differences (SB).

### 4.2. Series reactance

For this parameter, fewer studies show experimental results. Fig. 6 shows the comparison between the available measurements and simulation results, where armored and unarmored cables in both SP and SB are considered (no data available for armors with steel wires partially removed). The results for the reactance are obtained in the USM from the phase current and the total energy stored.

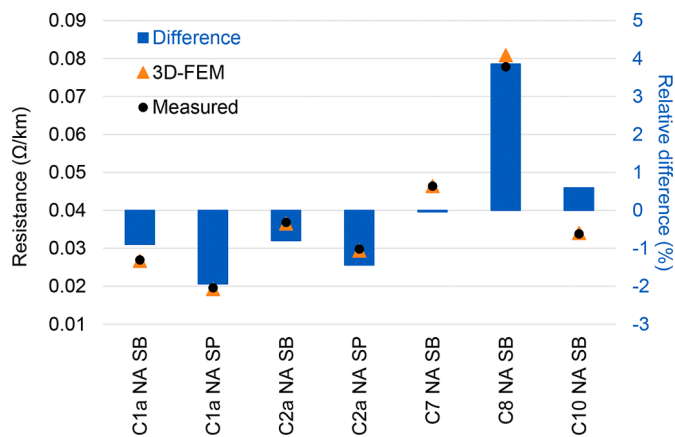

Fig. 3. Unarmored cables (SB and SP): measured and calculated series resistance (markers) and relative differences (bars).

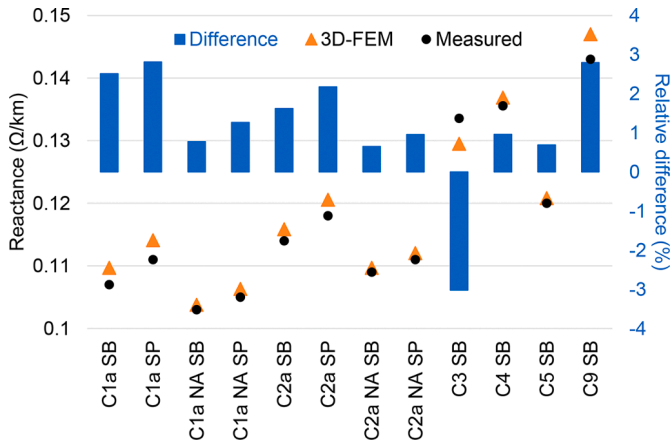

**Fig. 6.** SP and SB cables: measured and calculated series reactance (markers) and relative differences (bars).

It is to be noticed that different types of armor permeability are also considered in this study, having real or complex values, as well as nonlinear properties. In any case, the results show good agreement between measured and computed values, with relative differences of less than 3% in all the cases studied.

*4.3. Induced sheath current*

The induced sheath current when TCACs are in SB configuration is also a relevant topic to be evaluated by means of experimental setups. In this sense, Fig. 7 shows the experimental results found in the literature and 3D-FEM simulations for some of the studied cases. It is easily observed that there is usually a very good agreement between measurements and computed values, except for the smallest cable (C9), where the small values for the induced sheath current during tests (in the order of mA) may be affected by measurement tolerances. In most of the cases the relative difference is always below 4%, either for fully/half-armored or unarmored cables, and for linear or nonlinear steel properties. Moreover, this good agreement continues even when increasing the phase current, a situation where the nonlinear properties of the armor wires may play a relevant role. Thus, Fig. 8a shows how the induced sheath current increases with the current supplied during the tests in both, the experimental setups and the numerical simulations, with relative differences frequently below 6%, except for cable C9 (Fig. 8b).

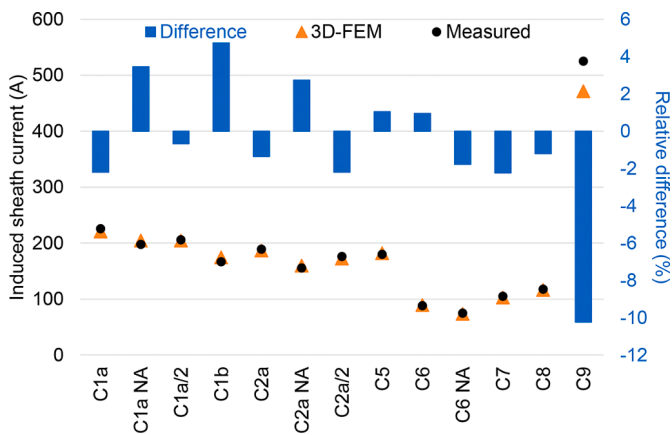

**Fig. 7.** SB cables: measured and calculated sheath current (markers) and relative differences (bars).

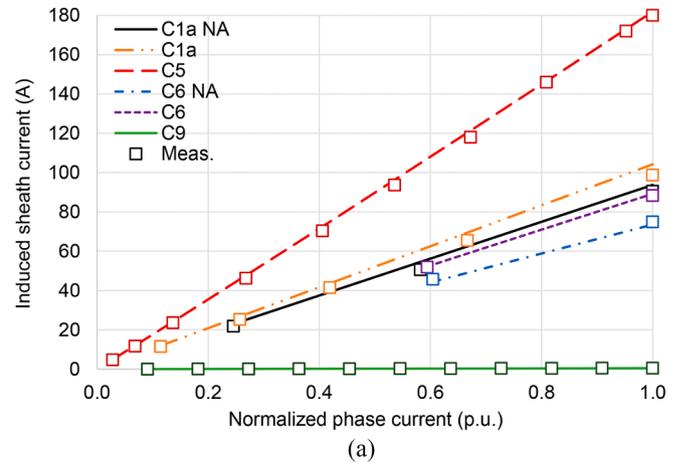

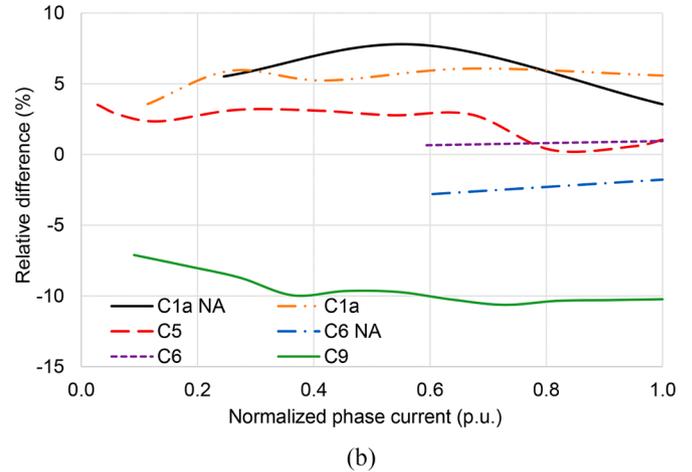

**Fig. 8.** SB cables: (a) Evolution of the measured and calculated sheath current with phase current and (b) relative differences.

*4.4. Zero sequence impedance*

Most of the experimental studies from the literature assume that the three-phase current is balanced. Even though the behavior of the USM with unbalanced current deserves a future in-depth analysis, we present the comparison results regarding the zero sequence impedance for two cases with experimental data: a 30 km, 150 kV TCAC that links Capri island with the Italian national transmission grid (cable C3) [37] and a 100 km, 220 kV TCAC that links Sicily and Malta (cable C4) [36,38]. For this task, the same current is injected through the three conductors, assuming the current return through the sheaths, armor and the surrounding medium. The latter is considered as homogeneous and with the properties of the sea, whose size has to be adequately extended for a proper evaluation of $R_0$ and $X_0$. Thus, following [48] the "infinite sea model" is assumed, so that the radius of the surrounding medium is taken equal to the Carson's depth:

$$d_{sea} \cong 503\sqrt{\frac{\rho_{sea}}{f}}, \qquad (3)$$

where $\rho_{sea}$ is taken as 200 mΩ·m [38,48]. This leads to $d_{sea} \cong$ 30 m in the 3D-FEM model, so the "infinite element domain" feature included in COMSOL is employed for scaling $d_{sea}$ in a shorter distance.

Table 2 summarizes the measurements reported in the literature, the results derived from 3D-FEM simulations and their relative differences. As can be observed, results are remarkable, with relative differences below 6 % despite the tolerances in measurements and the input data for the simulations.

**Table 2**
Cable C3 (SB): measured, calculated and relative difference for the zero sequence impedance.

|  | $R_0$ (Ω/km) | | | $X_0$ (Ω/km) | | |
| --- | --- | --- | --- | --- | --- | --- |
|  | Meas. | FEM | Diff. (%) | Meas. | FEM | Diff. (%) |
| C3 SB | 0.1844 | 0.1902 | 3.13 | 0.1475 | 0.1410 | −4.37 |
| C4 SB | 0.1545 | 0.1629 | 5.43 | 0.1309 | 0.1330 | 1.57 |

*4.5. Magnetic field distribution*

Few studies have been published regarding the MF distribution generated by TCACs. In this sense, Fig. 9 shows the results obtained in [24], where it is observed the evolution, along the cable length, of $B_l$, $B_v$ and $B_h$ over the surface of cable C10 (Fig. 10) when the armor is removed. Fig. 9 also includes the estimations derived from 3D-FEM simulations. Despite the possible uncertainties in the input data and measurements, there is a good agreement in the results, especially in $B_h$ and $B_v$.

From the environmental point of view it is also interesting to evaluate the MF distribution around the armored cable. In this sense, the authors of this work had the opportunity of taking MF measurements on cable C5 during the tests presented in [6]. These measurements were taken at different distances from the cable axis and at different heights from the ground (measurement lines ML1 and ML2 in Fig. 10), above which the TCAC was suspended at 1.24 m. During the tests a 3-axis EMDEX II MF meter was employed (resolution of 0.01 μT in the range of 0.01 to 300 μT at power frequency).

Having all this in mind, the measured and computed values in ML1 and ML2, as well as their relative differences, are shown in Figs. 11 and 12 for sheaths in SP and SB, respectively (logarithmic scale employed in the MF axis for better visualization).

As can be observed, in both situations differences are typically below 20 %, being the lowest discrepancies closer to the power cable, although they increase with the horizontal distance, especially in the SB case. These results may be probably caused by a combination of different factors, such as uncertainties in measurements, lower accuracy of the MF meter at low MF values or the small imbalance observed in the phase currents [6].

*4.6. Effects of the nonlinear permeability of the armor wires*

As commented earlier, the USM can be also employed in time-domain simulations for the analysis of different aspects that cannot be afforded through frequency-domain analyses. An example can be found in [6] for cable C5, where it is reported that the nonlinear properties of the armor wires give rise to a third harmonic in the induced armor voltage. To show this, different simulations have been performed for cable C5 considering linear ($\mu_r = 300$) and nonlinear properties (LG steel) for the steel wires. The results are represented in Fig. 13, where it is observed how, under balanced conditions, there is no induced voltage in the armor, as expected due to the relative twisting between phase conductors and the armor wires. Conversely, for LG steel wires, a 150 Hz waveform is obtained for the induced armor voltage, and still having a zero net value. Moreover, if there is a small imbalance in the phase currents, it can be seen how the linear steel gives rise to a 50 Hz voltage waveform, while the LG steel results in an induced voltage that contains both, the 50 Hz and 150 Hz components.

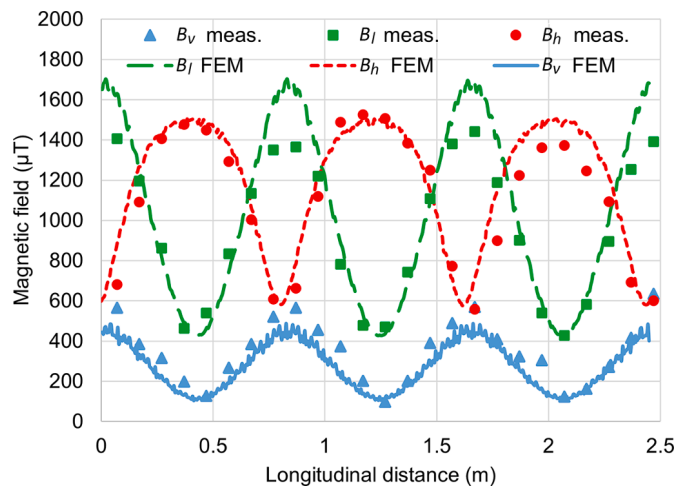

**Fig. 9.** C10 unarmored cable (SB): measured and calculated MF over cable C10 (Fig. 10) for 800 A of phase current.

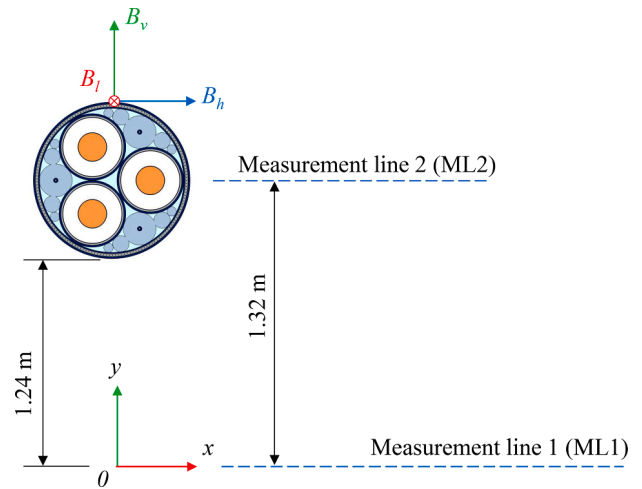

**Fig. 10.** MF: measurement point above cable C10 and measurement lines around cable C5.

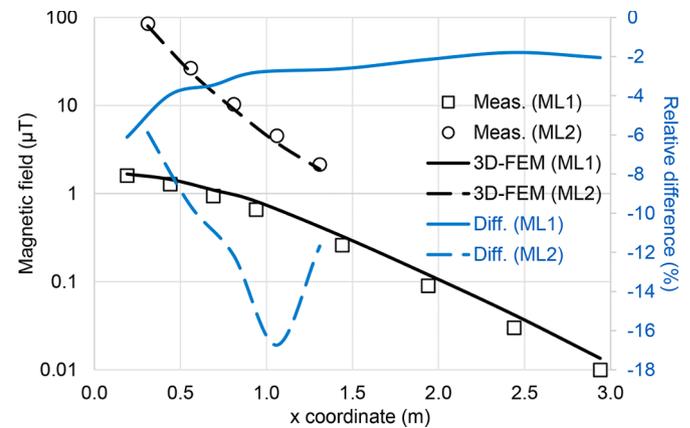

**Fig. 11.** SP: Evolution of the measured and computed MF with the distance at different heights and relative difference in cable C5 for 745 A.

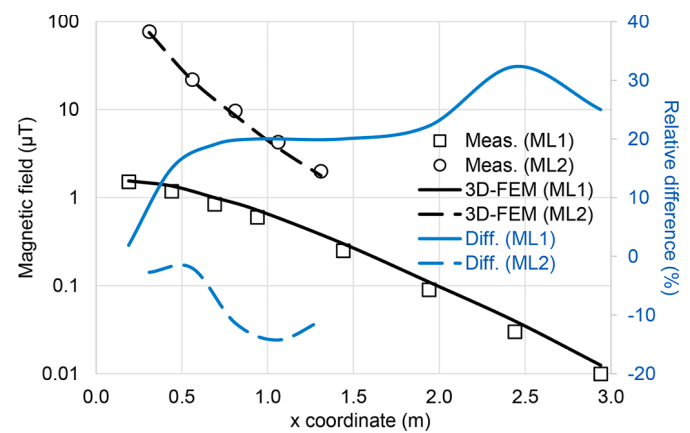

**Fig. 12.** SB: Evolution of the measured and computed MF with the distance at different heights and relative difference in cable C5 for 745 A.

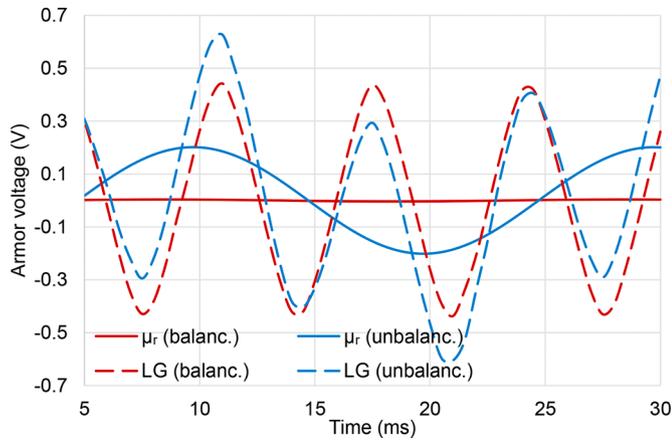

**Fig. 13.** Cable C5 (SB): induced armor voltage as a function of armor wire permeability.

These results are not only qualitatively in good agreement with those reported in [6], but also quantitatively. In this sense, [6] observed that the magnitude of this third harmonic increases nonlinearly with the phase current. This is shown in Fig. 14, where it can be observed a great match between the measurements reported by [6] and the results derived from 3D-FEM simulations, with relative differences below 7 %.

Finally, due to a small imbalance in the phase currents, a small net current was observed during the tests developed in [6]. Although there is no information regarding this imbalance and the value of the armor current, it has been simulated the case where the phase currents are 738∠1.4° A, 735∠ − 120° A and 742∠120.1° A. The results for $V_a$ and $I_a$ are depicted in Fig. 15 together with the measured values. Note that $I_a$ is shown as a relative waveform in p.u., since measurements are only available through its waveform in an oscilloscope snapshot represented in volts (Fig. 7 in [6]), with no additional information regarding its magnitude in amperes. Consequently, measurements and simulation values for $I_a$ can be only qualitatively compared in terms of its waveform.

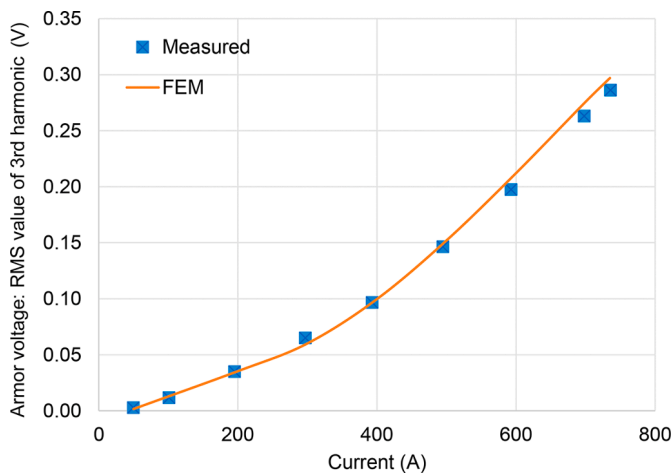

**Fig. 14.** Cable C5 (LG and SB): measured and computed armor voltage third harmonic as a function of the phase current.

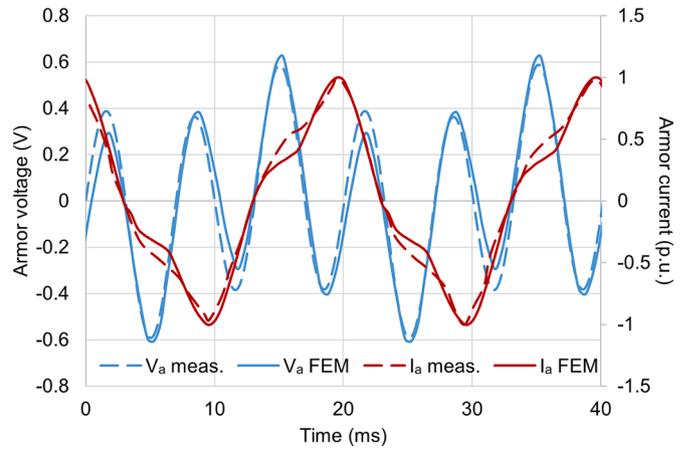

**Fig. 15.** Cable C5 (LG and SB): measured and computed armor voltage and current under unbalanced conditions.

As can be observed, there are small differences between the measured and computed voltages, both showing a similar amplitude in the fundamental and the third harmonic. Additionally, there is also a great similarity in the experimental and computed waveforms obtained for the armor current, despite the fact that they can be only compared qualitatively. In any case, these results highlight the good performance of the USM even for the analysis of intrinsic behaviors that take place in TCACs.

## 5. Conclusions

This paper presents the results of the experimental validation of the ultra-shortened 3D-FEM model for its application to the electromagnetic analysis of three-core armored cables at power frequency. To this aim, numerous experimental measurements for 10 real cables found in the literature have been used, including variants in bonding schemes and armor configurations. This validation is performed not only regarding classical parameters (e.g. series resistance), but also on aspects never addressed before through 3D-FEM simulations (e.g. zero sequence impedance).

The relative differences between measurements and computed values on estimating the series resistance, the inductive reactance and the induced sheath current are typically below 6%, either in cases with sheaths in SP and SB, as well as for unarmored, half-armored and fully-armored cables. Similar conclusions are obtained regarding the zero sequence impedance. Simulations also corroborate relevant aspects observed during experimental tests that are not addressed by the IEC 60287 standard, such as the influence of the phase current on the series resistance and the induced sheath current due to the nonlinear properties of the armor wires. Good results are also obtained when computing the MF distribution around three-core armored cables, although greater differences are observed than in the other parameters. Finally, time-domain simulations demonstrate the capability of the ultra-shortened 3D-FEM model to accurately reproduce intrinsic behaviors related to the nonlinear properties of the armor wires.

Thus, even considering the tolerances in measurements and typical uncertainties in the data employed in the simulations, remarkable results are obtained in this study, showing that the ultra-shortened 3D-FEM model is a suitable and trusted tool for designing and evaluating three-core armored cables, being a perfect complement to costly experimental setups. A future study will present a comparison of experimental tests and 3D-FEM simulations regarding electromagnetic parameters at harmonic frequencies.

**CRediT authorship contribution statement**

**Juan Carlos del-Pino-López:** Conceptualization, Methodology, Validation, Formal analysis, Investigation, Data curation, Writing – original draft. **Pedro Cruz-Romero:** Conceptualization, Formal analysis, Data curation, Writing – review & editing.

**Declaration of Competing Interest**

The authors declare that they have no known competing financial interests or personal relationships that could have appeared to influence the work reported in this paper.

**Acknowledgment**

This research was funded by FEDER/Ministerio de Ciencia e Innovación - Agencia Estatal de Investigación under the project ENE2017-89669-R and by the Universidad de Sevilla (VI PPIT-US) under grant 2018/00000740.

The authors would like to acknowledge and thank Jarle Bremnes and Marius Hatlo for bringing the opportunity to take MF measurements during their tests in the Halden factory.